\documentclass[10pt,a4paper,conference]{IEEEtran}
\usepackage{amsmath}
\usepackage{tikz}
\usepackage{tikz-inet}

\usepackage{pgf}
\usepackage{pgfplots}

\usepackage{amssymb,amsmath}
\usepackage{float}
\usetikzlibrary{shapes,snakes,arrows,automata}
\usepackage{algorithm}
\usepackage{algorithmic}
\usepackage{color}

\newtheorem{definition}{Definition}
\newtheorem{proposition}{Proposition}
\newtheorem{assumption}{Assumption}

\gdef\figwidth{8cm}
\gdef\figheight{6.3cm}

\newcommand{\rempat}[1]{\color{red}\footnote{\color{red}From Patrick: #1}\color{black}{}}

\renewcommand{\rempat}[1]{}

\newcommand{\WE}{^{\text{WE}}}
\newcommand{\opt}{^{\text{opt}}}
\newcommand{\obs}{}

\pgfcreateplotcyclelist{\mylist}{
	{black,mark=none,thin}, 
	{thick,densely dotted,mark=none},
	{densely dashed,mark=none},
	{dash pattern=on 8pt off 2 pt on 2pt off 2pt,thick,mark=none}, 
}

\pgfcreateplotcyclelist{\mylistcolor}{
	{blue,mark=none,thick,smooth}, 
	{red,mark=none,thick,smooth},
	{blue,mark=none,thick,densely dotted,smooth},
	{red,dash pattern=on 8pt off 2 pt on 2pt off 2pt,thick,mark=none,smooth}, 
}
\pgfcreateplotcyclelist{\mylistcolor3}{
	{blue,mark=none,thick}, 
	{red,mark=none,thick},
	{green,mark=none,thick},
	{red,dash pattern=on 8pt off 2 pt on 2pt off 2pt,thick,mark=none}, 
}
\pgfcreateplotcyclelist{\mylistmark}{
	{black,mark=o,thick}, 
	{black,mark=x,mark options=solid,thick,densely dashed},
	{black,mark=+,mark options=solid,thick,densely dotted},
	{mark=triangle,mark options=solid,dash pattern=on 8pt off 2 pt on 2pt off 2pt,thick}, 
}

\pgfcreateplotcyclelist{\mylistmarkcolor}{
	{blue,mark=o,thick}, 
	{blue,mark=x,mark options=solid,thick,densely dashed},
	{red,mark=+,mark options=solid,thick,densely dotted},
	{red,mark=triangle,mark options=solid,dash pattern=on 8pt off 2 pt on 2pt off 2pt,thick}, 
}

\pgfcreateplotcyclelist{\mylistcolor2}{
	{blue,mark=none, very thin}, 
	{red,mark=none,very thin},
	{green,mark=none,very thin},
	{black,very thin,mark=none}, 
}

\newcommand{\define}{:=}

\newcommand{\shortversion}[2]{#2}	

\begin{document}

\title{Efficiency or fairness: managing applications with different delay sensitivities in heterogeneous wireless networks}
\author{Vladimir Fux, Patrick Maill\'e, Jean-Marie Bonnin, Nassim Kaci\\Institut Mines-Telecom; Telecom Bretagne\\2, rue de la Chataigneraie CS 17607\\35576 Cesson-Sevigne, FRANCE\\
e-mail: \{first.last\}@telecom-bretagne.eu}


\maketitle

\begin{abstract}
In the current intensively changing technological environment, wireless network operators try to manage the increase of global traffic, optimizing the use of the available resources. 
This involves associating each user to one of its reachable wireless networks; a decision that can be made on the user side, in which case inefficiencies stem from user selfishness.

This paper aims at correcting that efficiency loss through the use of a one-dimensional incentive signal, interpreted as a price. While the so-called Pigovian taxes allow to deal with homogeneous users, we consider here two classes with different sensitivities to the Quality of Service, reflecting the dichotomy between delay-sensitive and delay-insensitive applications.
We consider a geographic area covered by two wireless networks, among which users 
choose based on a trade-off between the quality of service and the price to pay.

Using a non-atomic routing game model, we study analytically the case of constant demand levels. We show that when properly designed, the incentives elicit efficient user-network associations. Moreover, those optimal incentives can be simply computed by the wireless operator, using only some information that is easily available.
Finally, the performance of the incentive scheme under dynamic demand (users opening and closing connections over time) are investigated through simulations, our incentive scheme also yielding significant improvements in that case.
\end{abstract}

\section{Introduction}
The last years witnessed a tendency in the world of mobile devices, toward an increase in the number of different wireless network interfaces handled simultaneously. This gives mobile users a possibility to easily choose between several types of access points available, and soon it will be also possible that different applications on a mobile device use different network interfaces (each application favoring its best-suited network). 

Recently in~\cite{Gust}, the term \textit{always-best-connected} was coined, to describe a system where mobile devices automatically choose the most suitable network connection at every moment of time, depending on some concrete application needs, state of mobile device, economical reasons, etc. This kind of approach could cause a number of problems, one of which being the overload of some technologies and the underutilization of others. To cope with this particular problem, network providers have to use some sophisticated management schemes in order to avoid inefficient load distributions in their systems.

This topic is of high importance, and some recent papers address that same problem. A commonly used approach is to introduce a special bias for using lower-quality networks, which has a common interpretation as a tax imposed on higher-quality ones. How to determine appropriate taxes to influence users choices is discussed in \cite{Cole,Fleisher,Kara}.  We apply the same idea to a case of wireless heterogeneous networks, modeling the problem as a routing game.

The idea that users select the cheapest network with some additional penalty (tax) imposed by a network manager is not new. For the case when all users are equal from the point of view of tax perception (homogeneous users case), Beckmann et al.~\cite{Beck} showed that the so-called Pigovian taxes --applied on each link, and computed using the derivative of the cost functions of the links-- produce a minimum-latency (delay) traffic routing (see \cite{Pigou}). In~\cite{Cole} the case when users may perceive differently the relative costs of delay and taxes is considered (heterogeneous users case). It was proven that there also exist taxes which move a system to a situation when the average latency is minimized. These results have been generalized  to the multicommodity settings (i.e. several source-destination pairs) in~\cite{Kara,Kara09}.

While talking about effective users allocation, it is worth mentioning a totally different approach: to seek for an optimal user admission policy in the system through SMDP (Semi Markov Decision Processes). One could find this approach applied to the problem of global expected throughput maximization with the help of a central controller (taking admission decisions) in~\cite{Couch09,Coupe,Kumar}.

The contributions of this paper are as follows.
\begin{itemize}
\item Using the theoretical results of~\cite{Kara} proving the existence of taxes leading to an optimal flow repartition, we derive an analytical expression of that optimal tax.
\item We focus on the feasibility of the scheme, showing that the information needed to compute the optimal taxes is easily available to the network owner.
\item We evaluate, through simulations, the performance of our incentive mechanism in a more realistic context, where players are atomic and dynamically enter and leave the system over time.
\end{itemize}

\section{Model and Problem Formulation}
In this section, we present the network topology and the mathematical description of user behavior. More specifically, we investigate how users distribute over the different access networks, depending on their sensitivity, the taxes applied, and the QoS provided by the network access points. Finally, we summarize some previous results for the case without taxes, when users selfishly select wireless networks based on QoS.

\subsection{Network topology}
We consider a system with two wireless access networks (labeled by $1$ and $2$), owned by the same operator, who aims to achieve an efficient use of his access points. We assume that users seek an Internet connection through one of the two available networks, and their choices depend on the values of the taxes fixed by the operator and the QoS (here, the congestion-dependent latency) they experience. 

Two classes of users are considered, which are characterized by different sensitivities to the taxes. All users are assumed to have the choice between the two networks, i.e. they are in a common coverage area of both networks, and their devices support both networks' wireless access technologies. 

\subsection{Mathematical formulation}
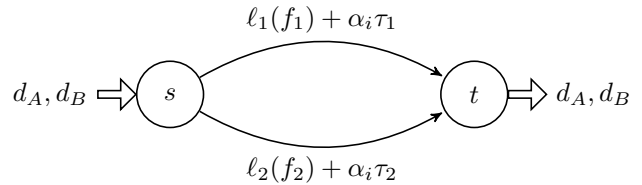
\begin{figure}[htbp]
\begin{center}
\begin{tikzpicture}[->,>=stealth',shorten >=1pt,auto,node distance=4cm,
                   semithick]
 \tikzstyle{every state}=[fill=white,draw=black,text=black]

 \node[state] (A)                    {$s$};
 \node[state]         (B) [right of=A] {$t$};

 \path (A) edge [bend right]             node[anchor=north] {$\ell_2(f_2) + \alpha_i \tau_2$} (B)
       (A) edge [bend left]             node {$\ell_1(f_1) + \alpha_i \tau_1$} (B);
\draw[-] (A.west) +(-.5,0) node (gauche){} +(-.5,-.05) -- +(-.2,-.05) -- +(-.2,-.2) -- +(0,0) -- +(-.2,.2) -- +(-.2,.05)  -- +(-.5,.05) -- cycle; %
\draw (gauche) node[anchor=east]{$d_A,d_B$}; 
\draw[-] (B.east) +(0,.05) -- +(.3,.05) -- +(.3,.2)  -- +(.5,0) node (droite){} -- +(.3,-.2) -- +(.3,-.05) -- +(0,-.05) -- cycle; %
\draw (droite) node[anchor=west]{$d_A,d_B$}; 
\end{tikzpicture}
\caption{Logic representation of the network selection problem as a routing problem.}
\label{fig:2arcs}
\end{center}
\end{figure}

From a logic point of view, the system can be seen as a routing problem with a directed graph composed of two parallel arcs, linking a source $s$ to a destination $t,$ as drawn on Figure~\ref{fig:2arcs}. The source node represents the (common) coverage area of networks $1$ and $2$, while the destination represents the worldwide Internet. Each one of the classes of users is assumed to be willing to send some flow to the Internet. To do so, users have to select one of the two access networks, modeled as two parallel links from $s$ to $t$.

We assume that users perceive the quality of a connection as a mixture of the QoS provided and the penalty (tax) imposed on the network. For sake of simplicity, we interpret the QoS value users are sensitive to as the latency (or delay) of connection. It is a matter of fact, that various types of applications suffer differently from delay degradation, e.g. for real-time voice and video applications admissible delay bounds are much smaller than for social network clients and web-browsing. For this reason, we consider a system where users perceive delay in different ways. We divide users into two classes based on their delay perception,  
so as to model the distinction between real-time and non-real-time traffic. Since users will have to weigh delay versus taxes when making their network choice, different delay perceptions correspond to different weights between tax and delay, or to different tax sensitivities if we keep the same weight for delay for both classes (our choice in this paper).

Each class $i$ of users, $i \in \{A,B\}$, is characterized in our model by an aggregate demand $d_i>0$ (the total throughput from class-$i$ users), and a tax-sensitivity parameter $\alpha_i$. We assume further, without loss of generality, that $\alpha_A > \alpha_B >0$, which means that class-$B$ users are more sensitive to delay than class-$A$ ones: if a tax is to be applied class-$B$ users focus more on delay, and are less likely to be influenced by the tax than class-$A$ users. 

An outcome of class $i$ users actions is a flow vector $f_i = (f_{1,i} , f_{2,i} )$, where  $f_{p,i}$  denotes a non-negative flow routed through network $p$. We call such flow vector feasible if $f_{1,i} +f_{2,i} = d_i$, i.e. if the sum of flows of class $i$ users routed through both networks is equal to the total demand of this class. We denote the total throughput demand of users by $D = d_A +d_B$, and assume it to be nonelastic (i.e. it does not depend on the taxes) and such that $D<c_1+c_2$. Also we will denote by $f = (f_1 , f_2)$ a feasible flow assignment of demand, i.e. a repartition of total demand $D = f_1 + f_2$ over two links, where $f_p \geq 0$ is the flow sent through network $p$.

The operator is assumed to be able to impose and modify tolls on its networks, that can be used to make revenue, but here mainly to drive the system to a more efficient resource usage. Similarly to~\cite{Kara}, we assume that the cost perceived by a class-$i$ user connected to network $p \in \{1, 2\}$ equals
\begin{eqnarray}
C_p(f)\define\ell_p(f_p) + \alpha_i \tau_p, &  i=1,2
\label{Eq:Cost}
\end{eqnarray} 
where $\tau_p$ is the tax imposed on network $p$. The delay on each network $p=1,2$ is assumed to increase with the network load $f_p$, through the delay function $\ell_p$ described below.

\begin{assumption}\label{ass:MM1}
Each network $p$ has a given capacity $c_p$, and without loss of generality, we assume that $c_2>c_1$.  The delay of a network carrying some flow level $f_p$ is assumed to be given by the mean sojourn time in an M/M/1 queue:
\begin{eqnarray}
\ell_p(f_p)= 
\begin{cases}
(c_p-f_p)^{-1} &\mbox{ if } f_p < c_p, \\
\; \infty &\mbox{ if }  f_p \geq c_p.
\end{cases}
\label{Eq:Assump}
\end{eqnarray}
\end{assumption}

The units used need to be clarified: modeling the packets as clients of an M/M/1 queue, the average sojourn time should be the one in~(\ref{Eq:Assump}), but multiplied by the packet size in the network. Assuming that the packet size is the same on both networks, we remove that multiplicative constant without loss of generality, leading to an interpretation of the tax $\tau_p$ as the price charged per packet sent on network $p$.

Note that we do not claim here that packet arrivals follow a Poisson process or that their size is exponentially distributed: we mainly choose the delay function~\eqref{Eq:Assump} because it reflects the congestion effects and is widely used in the literature. 


\subsection{Social cost}

Since we aim at optimally using the network, i.e. maximizing user satisfaction while satisfying demand, the global objective measure we consider here is the total delay $C(f)$, experienced by users. For a flow assignment $f=(f_1,f_2)$, this can simply be computed as
\begin{equation}
C(f) = f_1 \ell_1(f_1) + f_2 \ell_2(f_2),
\label{Eq:TotalCost}
\end{equation}
which is the total cost classically considered in routing games~\cite{Beck,stier2004selfish}.

\subsection{User Equilibrium}

In order to model user behavior, we follow a common assumption of users being selfish, in the sense that each user routes his flow to the network which minimizes his individual cost (delay+tax) given in~\eqref{Eq:Cost}. We assume that the number of users is large enough and all users have comparable demand levels, so that users are non-atomic \cite{Aum}, i.e. individual actions have no influence on the congestion levels. The cost functions given by~\eqref{Eq:Cost} define a game between users, where the equilibrium situation follows the so-called Wardrop's principle \cite{Ward}:
\begin{itemize}
\item{\textit{At equilibrium for each source-destination pair the travel costs on all the routes actually used are equal, or less than the travel costs on all non used routes.}}
\end{itemize}
Such a flow repartition is called a \textit{Wardrop equilibrium} among users. It is actually the non-atomic version of the more general concept of \textit{Nash equilibrium}~\cite{Fud}, i.e. a situation where no user has an interest in unilaterally changing his decision.

In our simple setting modeled by the graph on Figure~\ref{fig:2arcs}, a Wardrop equilibrium can be described mathematically as follows:

\begin{definition}
A Wardrop equilibrium is a feasible flow repartition $\bar{f} = (\bar{f_1}, \bar{f_2})$, decomposed into $\bar{f_p}=\bar{f}_{p,A}+\bar{f}_{p,B}$ for $p=1,2$, such that if $\bar{f}_{p,i}>0$ for some $p\in\{1,2\}$ and $i\in\{A,B\}$, then
\begin{equation}
\ell_p(\bar{f}_p) + \alpha_i \tau_p \leq \ell_{p'}(\bar{f}_{p'}) + \alpha_i \tau_{p'}
\label{Eq:Wardrop}
\end{equation}
where $p'\neq p$.
\end{definition}

In general, selfish user behavior leads to inefficient situations. By applying taxes, the operator aims to affect the selfish behavior outcome, eliciting them to spread more efficiently in the sense of the total cost~\eqref{Eq:TotalCost}.

\subsection{Previous analysis}\label{subsec:previous}
In this section we briefly recall some existing results for the routing problem considered, when no taxes are applied. 
Notice that without taxes, user heterogeneity in terms of QoS sensitivity does not have any impact, hence those results are independent of the decomposition of total demand $D$ into $d_A$ and $d_B$.
The proofs for the following results can be found in~\cite{Kaci09}.
\begin{itemize}
\item{Under Assumption~\ref{ass:MM1}, user selfish behavior leads to a unique Wardrop equilibrium $f\WE{}=(f_1\WE{},f_2\WE{})$:
\begin{eqnarray}
f\WE{}=
\begin{cases}
(0,D) \mbox{ if } D \leq c_2-c_1, \\
(\frac{D+c_1-c_2}{2}, \frac{D+ c_2 - c_1}{2}) \mbox{ otherwise.}
\end{cases}
\label{fig:WE}
\end{eqnarray}
}
\item{An optimal flow assignment $f\opt{} = (f_1\opt{},f_2\opt{})$ in the sense of the global cost defined in~\eqref{Eq:TotalCost} is the solution of the following mathematical program:
\begin{eqnarray}
\min_{f_1,f_2} \; f_1 \ell_1(f_1) + f_2 \ell_2(f_2) \\ \nonumber
\mbox{s.t. }   \;
\begin{cases}
f_1+f_2 = D \\
f_1 \geq 0, f_2 \geq 0
\end{cases}
\label{Eq:Minim}
\end{eqnarray}
Under Assumption~\ref{ass:MM1}, there exists a unique optimal assignment. Using Karush-Kuhn-Tucker optimality conditions \cite{Rock}, we have
\begin{eqnarray}
f\opt{}\!\! = \!\!
\left\{
\!\!\!\!
\begin{array}{lr}
(0, D) &  \hspace{-3.5cm} \mbox{ if } D \!\leq\! c_2 \!-\! \sqrt{c_1 c_2}, \\
\left( \frac{(D-c_2) \sqrt{c_1} + c_1 \sqrt{c_2}} {\sqrt{c_1} + \sqrt{c_2}} ,\!  \frac{(D-c_1) \sqrt{c_2} + c_2 \sqrt{c_1}} {\sqrt{c_1} + \sqrt{c_2}} \right) & \mbox{ oth.} ,
\end{array}
\right.
\label{Eq:OptFlows}
\end{eqnarray}

with the corresponding total cost 
\begin{eqnarray}
C\opt{} = 
\begin{cases}
\frac{D}{c_2-D} & \mbox{ if } D \leq c_2 - \sqrt{c_1 c_2}, \\
\frac{2D - c_1 - c_2 + 2 \sqrt{c_2 c_1}} {c_1+c_2 - D} & \mbox{ otherwise}.
\end{cases}
\label{Eq:OptCost}
\end{eqnarray}
}

\item Under Assumption~\ref{ass:MM1}, for given values of network capacities $(c_i)_{i=1,2}$ and total demand $D$, we have
$f_2\opt{} \leq f_2\WE{}$, i.e. the traffic flow which is routed through network $2$ (the largest capacity network) at equilibrium is greater than or equal to the optimum traffic flow in this network. Therefore the tax should be applied to network $2$, to elicit some users to rather use network $1$.

\item Under Assumption~\ref{ass:MM1}, for given values of network capacities $(c_i)_{i=1,2}$ and total demand $D$, 
the total cost is minimized at equilibrium when $D \leq c_2 - \sqrt{c_1 c_2}$, hence no tax is necessary in that case.
\end{itemize}
This paper extends the model studied in~\cite{Kaci09}, introducing a distinction between the classes of users expressed through their tax (or equivalently, QoS) sensitivities.

\section{Eliciting optimal user-network associations with taxes}

In this section we derive an analytical expression of the taxes yielding an optimal flow repartition, and discuss the information required for their computation. The existence of such optimal taxes has been established in~\cite{Cole,Fleisher,Kara}. 

\subsection{Expression of optimal taxes}
As we pointed out in Subsection~\ref{subsec:previous}, under Assumption~\ref{ass:MM1} a tax needs to be applied to the largest-capacity network (network $2$ in our model), and only when the system is sufficiently loaded. Let us denote by $\tau_2$ that optimal tax: when applied to network $2$, the Wardrop equilibrium $\bar f$ among users must be equal to the optimum traffic flow:
\begin{eqnarray}
\begin{cases}
\bar{f}_{1,A} + \bar{f}_{1,B} =  f_1\opt{} ,\\
\bar{f}_{2,A} + \bar{f}_{2,B} =  f_2\opt{}.
\end{cases}
\label{Eq:FlowsCon}
\end{eqnarray}

The following proposition shows that when the system is sufficiently loaded, the expression of the optimal tax depends only on total demand $D$, and on the sign of $(d_B - f_2\opt{})$.

\begin{proposition}\label{prop:opt_tax}
Under Assumption~\ref{ass:MM1}, for given values of network capacities $(c_p)_{p=1,2}$, demand $D=d_A+d_B<c_1+c_2$, and sensitivities $(\alpha_i)_{i \in \{A,B\}}$, an optimal tax $\tau_2$ to apply to network $2$ when $D > c_2 - \sqrt{c_1 c_2}$ is given by
\begin{eqnarray}
\tau_2= 
\begin{cases}
\frac{c_2-c_1} {\alpha_A \sqrt{c_1 c_2} ( c_2 + c_1 - D)} & \mbox{if } d_B \leq f_2\opt{},\\
\frac{c_2-c_1} {\alpha_B \sqrt{c_1 c_2} ( c_2 + c_1 - D)} & \text{otherwise}.
\end{cases}
\label{Eq:OptTax}
\end{eqnarray}
When $D \leq c_2 - \sqrt{c_1 c_2}$, no tax is necessary.
\end{proposition}

\shortversion{The proof is omitted due to lack of space, the interested reader is referred to~\cite{fux2013efficiency_hal}.}
{
\begin{IEEEproof}
Recall that we assumed without loss of generality that $ c_2 > c_1$ and $\alpha_A > \alpha_B$.
\par\noindent$\bullet$ We first treat the case when $d_B - f_2\opt{} \leq 0$. From the equality $d_B=\bar{f}_{1,B} + \bar{f}_{2,B}$ and the expression of $f_2\opt{}$ in~\eqref{Eq:FlowsCon}, we have
\[
\bar{f}_{1,B} + \bar{f}_{2,B} \leq \bar{f}_{2,A} + \bar{f}_{2,B}  \text{,\quad i.e., \quad}
\bar{f}_{1,B} \leq \bar{f}_{2,A} .
\]
Assume that $\bar{f}_{1,B} > 0$: then necessarily $\bar{f}_{2,A} >0$. We thus obtain, from~\eqref{Eq:Wardrop} and~\eqref{Eq:FlowsCon}:
%
\begin{equation}
\ell_1(f_1\opt{}) \leq \ell_{2}(f_2\opt{}) + \alpha_B \tau_{2}
\label{Eq:TaxProof1}
\end{equation}
and 
\begin{equation}
\ell_{2}(f_2\opt{}) + \alpha_A \tau_{2}\leq \ell_1(f_1\opt{}),
\label{Eq:TaxProof2}
\end{equation}
%
which yields $\alpha_A\leq \alpha_B$, a contradiction. As a consequence, we necessarily have $\bar{f}_{1,B} = 0$.


We now distinguish two cases. 
\begin{itemize}
\item[-] First, if $\bar{f}_{2,A} =0$, then~\eqref{Eq:FlowsCon} implies that $d_A=\bar{f}_{1,A} =f_1\opt{}$ and $d_B= \bar{f}_{2,B} =f_2\opt{} $. But  we have $f_1\opt{}>0$ and $f_2\opt{} > 0$ when $D > c_2 - \sqrt{c_1 c_2}$, which implies that $\bar{f}_{1,A} > 0 $ and $\bar{f}_{2,B} > 0 $. 
From~\eqref{Eq:Wardrop} and~\eqref{Eq:FlowsCon}, this implies
\begin{equation}
\ell_1(f_1\opt{}) \leq \ell_2(f_2\opt{}) + \tau_2 \cdot \alpha_A  
\label{Eq:TaxProof3}
\end{equation}
and
\begin{equation}
\ell_2(f_2\opt{}) + \tau_2 \cdot \alpha_B \leq \ell_1(f_1\opt{})
\label{Eq:TaxProof4}
\end{equation}
Since $\alpha_A>\alpha_B$, Inequalities~\eqref{Eq:TaxProof3} and~\eqref{Eq:TaxProof4} simultaneously hold if and only if 
\[
\frac{1}{\alpha_A}\!\left(\ell_1(f_1\opt{})-\ell_2(f_2\opt{})\!\right)\!\leq\! \tau_2 \!\leq\! \frac{1}{\alpha_B}\!\left(\ell_1(f_1\opt{})-\ell_2(f_2\opt{})\!\right)
\]
which can be rewritten, using our specific cost functions 
and
replacing $f_1\opt{}$ and $f_2\opt{}$ with their values given in~\eqref{Eq:OptFlows},
\[
\frac{c_2 - c_1} {\alpha_A \sqrt{c_1 c_2} (c_2 + c_1 - D)}\leq\tau_2\leq \frac{c_2 - c_1} {\alpha_B \sqrt{c_1 c_2} (c_2 + c_1 - D)},
\]
which is satisfied by the form proposed in~(\ref{Eq:OptTax}).

\item[-] Now suppose that $\bar{f}_{2,A}>0$ (recall that $\bar{f}_{1,B} =0$), then $\bar{f}_{1,A} >0$, $\bar{f}_{2,B} >0$  and $\bar{f}_{2,A} >0$. Relation~\eqref{Eq:TaxProof4} should be verified and~\eqref{Eq:TaxProof3} must hold \emph{with equality}, which is the case if and only if
\begin{equation}
\tau_2 = \frac{c_2 - c_1} {\alpha_A \sqrt{c_1 c_2} (c_2 + c_1 - D)}.
\end{equation}
\end{itemize}
\par\noindent $\bullet$ We now treat the case when $d_B - f_2\opt{} > 0$. From~(\ref{Eq:FlowsCon}) we obtain
\begin{eqnarray*}
\bar{f}_{1,B} + \bar{f}_{2,B} > \bar{f}_{2,A} + \bar{f}_{2,B}  
\text{, \quad  i.e. \quad}
\bar{f}_{1,B} > \bar{f}_{2,A} \nonumber.
\end{eqnarray*}

Thus, $\bar{f}_{1,B} > 0$. But we have shown in the previous case that we cannot simultaneously have $\bar{f}_{2,A} > 0$, thus $\bar{f}_{2,A} = 0$.
On the other hand we have $\bar{f}_{2,B} > 0$ from~\eqref{Eq:FlowsCon} and~(\ref{Eq:OptFlows}), since $f_2\opt>0$
. The inequality~\eqref{Eq:TaxProof2} therefore holds, as well as~\eqref{Eq:TaxProof1} that must hold \emph{with equality}. This is the case if and only if
%
\begin{equation*}
\tau_2 = \frac{c_2 - c_1} {\alpha_B \sqrt{c_1 c_2} (c_2 + c_1 - D)},
\end{equation*}
which establishes the proposition for the case when $D > c_2 - \sqrt{c_1 c_2}$. When $D \leq c_2 - \sqrt{c_1 c_2}$, no tax is needed as evidenced in Subsection~\ref{subsec:previous}.
\end{IEEEproof}
}

\subsection{Effect of optimal taxes on the perceived delay}

\begin{figure}[htbp]
{\footnotesize
\begin{tikzpicture}
\begin{axis}[
width=\figwidth,height=\figheight,
xlabel=$d_A/D$,
ylabel={Average latency},
every axis legend/.append style={nodes={right}}, 
]
\addplot[thin,domain=0:1] {
min(0.379,(0.516 + (6.644 - 8 * (1-x))/(11-6.644) ) / (8 * x))
 };
\addlegendentry{Class-$A$ users}
\addplot[very thick,domain=0:1,samples=200] {
max((6.644*0.224+ (1.366 - 8 * x) * 0.378) /  (8 * (1-x) ),0.224)
};
\addlegendentry{Class-$B$ users}
\addplot[thick,dotted,domain=0:1] {
0.286
 };
\addlegendentry{No-tax case}
\draw[dashed] (axis cs:0.171,\pgfkeysvalueof{/pgfplots/ymin})--node[near end,rotate=90,anchor=south]{$f_1\opt{}\!/\!D$} (axis cs:0.171,\pgfkeysvalueof{/pgfplots/ymax});
\end{axis}
\end{tikzpicture}
} 
\caption{Average latency of both classes at the no-tax equilibrium and with optimal taxes. In the latter case, when $d_A<f_1\opt{}$ (resp. $d_A>f_1\opt{}$) all type-$A$ (resp. type-$B$) users select network $1$ (resp. $2$). Parameter values: $D=8$Mbit/s, $c_1=4$Mbit/s, $c_2=11$Mbit/s.}
\label{fig:ClassLat}
\end{figure}
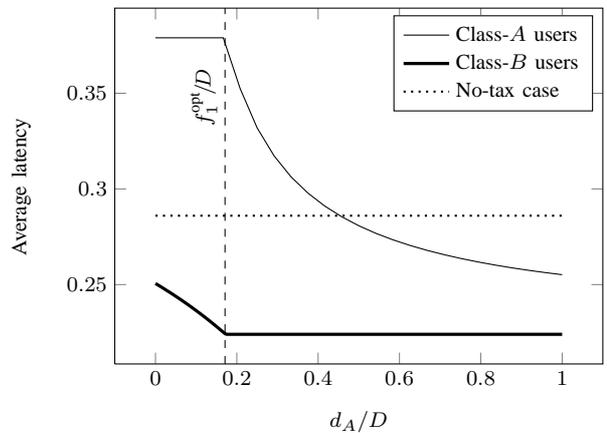

As intended, our approach allows to separate delay-sensitive and delay-insensitive users, as illustrated on Figure~\ref{fig:ClassLat}, where the average latencies experienced by both user classes are plotted when the proportion of class-$A$ users vary, and compared to the no-tax case (where both classes have the same latency). Note that delay-sensitive class-$B$ users benefit from the best quality in average (lower latency), at the expense of class-$A$ users which suffer higher delay but are less sensitive to it.

\subsection{Practical issues} 

The total demand $d_B$ of users of class $B$ can be hard to measure in practice, since the network owner can not determine the exact number of users from each class being connected to his networks at a particular moment of time (and users may have incentives to misreport their types if asked to declare it). However, we recall that the exact value of $d_B$ is not needed:  Indeed, both expressions of the optimal tax in~(\ref{Eq:OptTax}) only involve the total demand $D$, that is directly observable by the network owner. Only the \emph{sign} of $d_B-f_2\opt$ has to be determined to select the appropriate form. An approximation we propose is to use the average load of users of class $B$, which is much easier to determine, as an estimator of $d_B$. We can indeed assume that the arrival process of class $B$ members and the time they spend in average in the network are known stochastically, and calculate the average class $B$ load. The impact of such an approximation will be evaluated through simulations in the next section.

\section{Simulation Scenario}
This section complements the mathematical analysis, by providing a simulation model to evaluate the performance of our tax mechanism in a wireless network where users dynamically enter and leave the system.

\subsection{Simulation model and scenarios}

We consider a simple scenario where the operator owns two access points, with respective WiFi implementations IEEE 802.11b ($c_1 = 4$Mbit/s) and IEEE 802.11g ($c_2 = 11$Mbit/s). We assume  mobile users of class $i \in \{A, B\}$ join the system according to a Poisson process with parameter $\lambda_i$. We further assume that the classes correspond to different services:
\begin{itemize}
\item Streaming audio (non-real time: music or radio, for example) for users of class $A$, with individual throughput $\epsilon_A = 0.064$Mbit/s.
\item Delay-sensitive (real-time) video conversation call for users of class $B$, with individual throughput $\epsilon_B = 0.184$Mbit/s. 
\end{itemize}
Those definitions are compliant with our model convention, where type-$B$ users are more delay-sensitive than type-$A$, thus $\alpha_A>\alpha_B$.

Note that each user has a non-zero individual throughput, hence the game is not perfectly non-atomic. Nevertheless, the individual throughput values are small with regard to the network capacities ($c_1 = 4$Mbit/s, $c_2 = 11$Mbit/s), so that the impact on QoS of individual choices remain small.

Each user is connected to the network for a duration modeled as a random variable, following an exponential distribution with parameter $\mu_i$. The average listening time of class-$A$ users is therefore $\frac{1}{\mu_A}$ (seconds), and the average video conversation time of class-$B$ users is $\frac{1}{\mu_B}$ (seconds). Users choose an access network upon their arrival, selecting the cheapest one in the sense of their cost (given in Equation~\eqref{Eq:Cost}). 
We investigate two settings: in the first setting, users remain attached to the same network for the whole duration of their connection (no handovers), even if QoS conditions vary. In the second setting, vertical handovers between networks can occur.

Note that under our assumptions, the process describing the number of users of each class in each network is a continuous-time Markov chain, which we study through simulations due to the excessively large number of states.
Note that the latency function we consider (Equation~\eqref{Eq:Assump}) are only defined when demand is below capacity. We tackle this problem by dropping  the arriving users for which there is no sufficient available capacity on any network. The resulting blocking rate is measured in our simulations.

We investigate three different scenarios for each aforementioned simulation setting. In the first scenario, the tax is not applied at all - the users act without any intervention from the provider's side. In the second scenario, the operator is willing to apply the optimal tax expressed in Proposition~\ref{prop:opt_tax}, but is not able to measure the exact value of $d_B$. In that case the tax expression is chosen based on the \emph{average load} of class-$B$ users $(\hat d_B =  \frac{\epsilon_B  \lambda_B }{\mu_B} )$ in the network, assuming that the arrival rate ${\lambda_B}$ of type-$B$ users and their average connection duration $\frac{1}{\mu_B}$ are  known by the operator. The third scenario, called the optimum situation, assumes that the operator is able to determine precisely the load $d_B$ of class-$B$ users, and thus to apply the exact optimal tax of Proposition~\ref{prop:opt_tax}. Recall that for the scenarios involving taxes, the tax is applied only when the network load $D$ exceeds $c_2 - \sqrt{c_1 c_2}$, i.e. when selfish user behavior does not lead to an optimal situation.

\subsection{Simulation results}

This section presents the results obtained with the simulation scenarios described above, for the parameter values
 $\frac{1}{\mu_A} = 4$ minutes, $\frac{1}{\mu_B} = 2.5$ minutes, $\lambda_A = 3$ (arrivals/minute), $\lambda_B =4.5$ (arrivals/minute), and the tax sensitivities parameters $\alpha_A=2$ and $\alpha_B=1$ (cost units per (dollar per packet)).

In particular, we analyze the (average) Price of Anarchy (PoA), that is the ratio between the total delay (Equation~\eqref{Eq:TotalCost}) resulting from selfish user behavior and the minimum total delay~\cite{Kara}. This metric helps us investigate the efficiency of tax application for different flow conditions. We finally compare both settings, whether vertical handovers can occur or not.

The evolution of the total network load for one simulation trajectory is shown on Figure~\ref{fig:TotalLoad}, with the horizontal line corresponding to the threshold value of total load $c_2 - \sqrt{c_1 c_2}$, above which taxes are needed to reduce the total delay $C$. 
\gdef\thexmin{10}
\gdef\thexmax{16}
\begin{figure}[htbp]
{\footnotesize
\begin{tikzpicture}
\tikzset{
every pin/.style={fill=yellow!50!white,rectangle,rounded corners=3pt,font=\tiny},
small dot/.style={fill=black,circle,scale=0.3}
}
\begin{axis}[
width=\figwidth,height=\figheight,
xmin=\thexmin,
xmax=\thexmax,
xlabel={Time (min) },
ylabel={Total Load [Mbit/s]},
cycle list name=\mylist,
every axis legend/.append style={nodes={right}}, 
]
\addplot plot[const plot] file {totalLoadF.dat};
\addplot coordinates {
(0, 4.36) (20, 4.36) 
};
\node[small dot,pin=120:{$(0.5,0,5)$}] at (axis description cs:0,0) {};
\draw (axis cs:15, 4.36) node[anchor= south]{$c_2 - \sqrt{c_2 c_1}$};
\end{axis}
\end{tikzpicture}
} 
\caption{Total load versus time}
\label{fig:TotalLoad}
\end{figure}
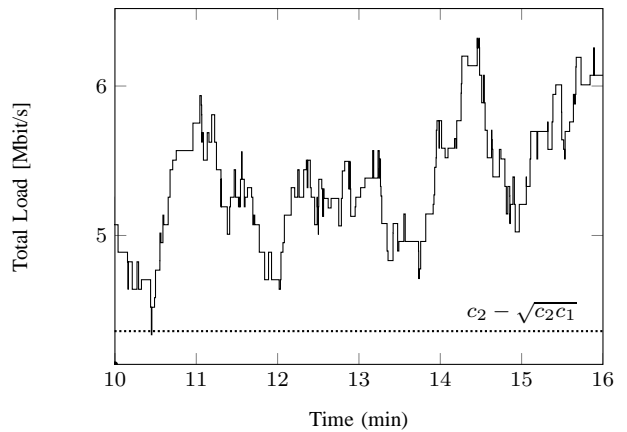
\gdef\notaxcase{No tax}
\gdef\approxtax{Approx. tax}
\gdef\optimaltax{Optimal tax}

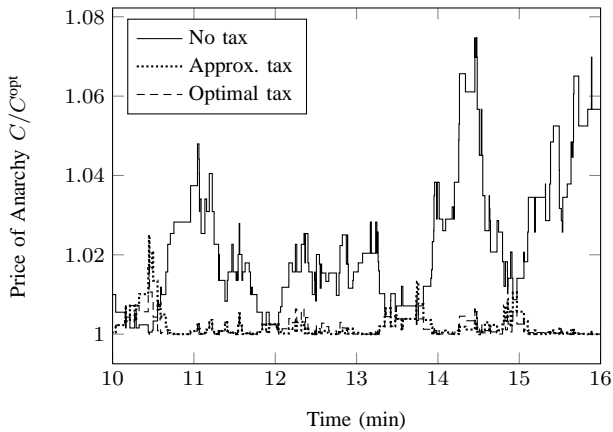
\begin{figure}[htbp]
{\footnotesize
\begin{tikzpicture}
\begin{axis}[
width=\figwidth,height=\figheight,
xmin=\thexmin,
xmax=\thexmax,
xlabel={Time (min)},
ylabel={Price of Anarchy $C\obs/C\opt$},
cycle list name=\mylist,
legend style={at={(0.03,0.97)},anchor=north west},
every axis legend/.append style={nodes={right}}, 
]
\addplot plot[const plot] file {noTaxEfficiencyF.dat};
\addlegendentry{\notaxcase}
\addplot plot[const plot] file {withTaxEfficiencyF.dat};
\addlegendentry{\approxtax}
\addplot plot[const plot] file {optimalTaxEfficiencyF.dat};
\addlegendentry{\optimaltax}
\end{axis}
\end{tikzpicture}
} 
\caption{Price of Anarchy versus time, without vertical handovers.}
\label{fig:PoA}
\end{figure}

Figure~\ref{fig:PoA} displays the corresponding evolution over time of the Price of Anarchy (PoA) for the three tax scenarios, when no vertical handovers occur  (i.e., users do not constantly adapt to the changes in delay and price).
We notice 
that even in that case, taxation can yield significant performance gains.
Interestingly, we remark that the optimal tax does not always imply the lowest total delay, the total delay with that tax being sometimes even higher than without any tax. We have to recall here that there are several differences between our mathematical model and the simulation model considered in this section. Notably, we do not allow here users to switch networks, which can lead to the following situation. Consider some moment of time when the total flow suddenly falls below the $c_2 - \sqrt{c_1 c_2}$ threshold; the optimal flow in the network $1$ then equals zero (see~\eqref{Eq:OptFlows}). But in general more users (among those still in the system) had chosen network $1$ when a tax was previously applied on network $2$, hence the no-tax case leads temporarily to a situation closer to the optimal one. In other words, our simulation system without vertical handovers shows some inertia: the flow distribution cannot instantly change when QoS conditions evolve. This situation occurs in Figure~\ref{fig:PoA} around $t=10$ minutes for example, and similar cases (when demand suddenly drops and inertia impacts the outcome) occur around time $t=14$ and $t=15$ minutes even if demand remains above $c_2 - \sqrt{c_1 c_2}$. 

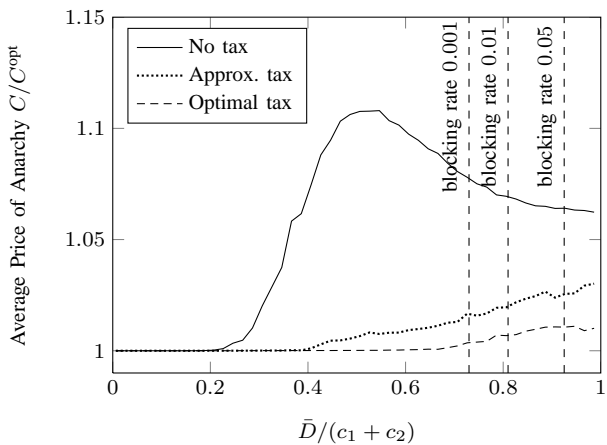
\begin{figure}[htbp]
{\footnotesize\begin{tikzpicture}
\begin{axis}[
width=\figwidth,height=\figheight,xmin=0,ymax=1.13,ymin=0.99,xmax=1,
xlabel={ $\bar{D}/ (c_1 + c_2)$},
ylabel={Average Price of Anarchy $C\obs/C\opt$},
ymax=1.15,
cycle list name=\mylist,
legend style={at={(0.03,0.97)},anchor=north west},
every axis legend/.append style={nodes={right}}, 
]
\addplot plot file {noTaxAverageCost1_5SwitchesF.dat};
\addlegendentry{\notaxcase}
\addplot plot file {withTaxAverageCost1_5SwitchesF.dat};
\addlegendentry{\approxtax}
\addplot plot file {optimalTaxAverageCost1_5SwitchesF.dat};
\addlegendentry{\optimaltax}
\draw[dashed] (axis cs:0.73,\pgfkeysvalueof{/pgfplots/ymin})--node[near end,rotate=90,anchor=south]{blocking rate $0.001$} (axis cs:.73,\pgfkeysvalueof{/pgfplots/ymax});
\draw[dashed] (axis cs:0.81,\pgfkeysvalueof{/pgfplots/ymin})--node[near end,rotate=90,anchor=south]{blocking rate $0.01$} (axis cs:0.81,\pgfkeysvalueof{/pgfplots/ymax});
\draw[dashed] (axis cs:0.925,\pgfkeysvalueof{/pgfplots/ymin})--node[near end,rotate=90,anchor=south]{blocking rate $0.05$} (axis cs:0.925,\pgfkeysvalueof{/pgfplots/ymax});
\end{axis}
\end{tikzpicture}
} 
\caption{Average Price of Anarchy versus load with vertical handovers.}
\label{fig:PoA15Sw}
\end{figure}

\begin{figure}[htbp]
{\footnotesize\begin{tikzpicture}
\begin{axis}[
width=\figwidth,height=\figheight,xmin=0,ymax=1.08,ymin=0.99,xmax=1,
xlabel={ $\bar{D}/ (c_1 + c_2)$},
ylabel={Average Price of Anarchy $C\obs/C\opt$},
ymax=1.10,
cycle list name=\mylist,
legend style={at={(0.03,0.97)},anchor=north west},
every axis legend/.append style={nodes={right}}, 
]
\addplot plot file {noTaxAverageCost1_5NoSwF.dat};
\addlegendentry{\notaxcase}
\addplot plot file {withTaxAverageCost1_5NoSwF.dat};
\addlegendentry{\approxtax}
\addplot plot file {optimalTaxAverageCost1_5NoSwF.dat};
\addlegendentry{\optimaltax}
\draw[dashed] (axis cs:0.70,\pgfkeysvalueof{/pgfplots/ymin})--node[near end,rotate=90,anchor=south]{blocking rate $0.001$} (axis cs:.70,\pgfkeysvalueof{/pgfplots/ymax});
\draw[dashed] (axis cs:0.80,\pgfkeysvalueof{/pgfplots/ymin})--node[near end,rotate=90,anchor=south]{blocking rate $0.01$} (axis cs:0.80,\pgfkeysvalueof{/pgfplots/ymax});
\draw[dashed] (axis cs:0.91,\pgfkeysvalueof{/pgfplots/ymin})--node[near end,rotate=90,anchor=south]{blocking rate $0.05$} (axis cs:0.91,\pgfkeysvalueof{/pgfplots/ymax});
\end{axis}
\end{tikzpicture}
} 
\caption{Average Price of Anarchy versus load without vertical handovers.}
\label{fig:PoA15NoSw}
\end{figure}
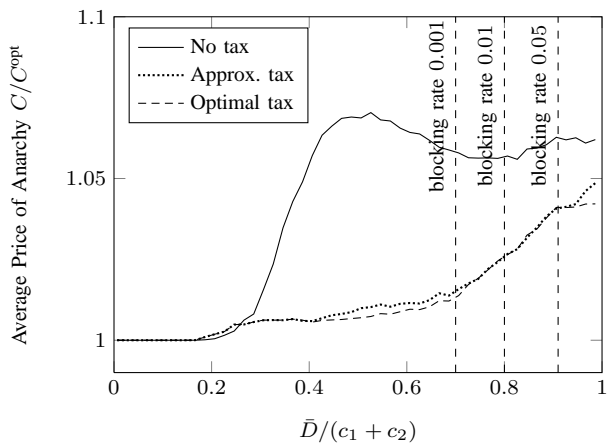

Those phenomena being highlighted on one trajectory, we now turn our attention to their statistical impact, through extensively many simulations of the same scenarios.
The results of these repeated simulations are presented on Figures~\ref{fig:PoA15Sw}-\ref{fig:PoA15NoSw}, when the total average load $\frac{\bar D}{c_1+c_2}$ varies, with $\bar D=\frac{\lambda_A}{\mu_A}\epsilon_A+\frac{\lambda_B}{\mu_B}\epsilon_B$:the parameters evolving are $\lambda_A$ and $\lambda_B$, with their ratio maintained at $\lambda_A/\lambda_B=1.5$.


We first notice from Figures~\ref{fig:PoA15Sw}-\ref{fig:PoA15NoSw} that the no-tax curve has a form similar to the one predicted by the theoretical study in~\cite{Kaci09}. On those figures, we also depicted the demand thresholds corresponding to some blocking rates values (proportion of users rejected due to lack of capacity). Since wireless systems are designed to have low blocking rates (below $1 \%$), those values show that the load range of interest lies between $0$ and $0.8$. 

When vertical handovers are permitted, we observe that optimal taxes drive the system very close to the optimal situation - the Price of Anarchy (PoA) does not rise above $1.01$. For the approximate tax the PoA can reach $1.03$, which is still significantly lower than the PoA of the no-tax case (that goes up to $1.10$). We also observe a significative influence on efficiency, of the presence of vertical handovers. As mentioned before, prohibiting handovers prevents the system from balancing rapidly the load among networks, implying larger costs.

A curious phenomenon worth mentioning from Figure~\ref{fig:PoA15NoSw} is the small range of average total load for which the average PoA of the no-tax case is lower than for the case with taxes (load values between $0.2$ and $0.3$). This suggests that the situations explained before on the single trajectory (Figure \ref{fig:PoA}) are not so rare in that case.
Indeed, as a result of total load being low, at very few moments of time does the load exceed $c_2 - \sqrt{c_1 c_2}$, which causes the taxes introduction, deterring new entrants from using network $2$. But this situation does not hold for a long time: quite soon the load goes below the threshold value, and because of handovers being forbidden, the flow in the first network remains positive, causing inefficiencies. Nevertheless, this inefficiency range remains small, and the PoA difference is limited, so this does not question the gain of our incentive mechanism. In our simulations, the taxation approach appears to be most effective for average total loads above $30\%$ (for the considered simulation parameters) of the total capacity, and the highest efficiency gain is reached around loads corresponding to $50\%$ of the total capacity.

\section{Conclusion and future work}

This paper focuses on the selfish behavior of users in wireless systems, and on the possibility of influencing it through taxation in the particular case when delay-sensitive and delay-insensitive flows coexist. We have derived the analytical expression for the optimal tax in a wireless heterogeneous network scenario. Simulation results highlight a significant performance gain, both for the situation when the operator has a complete, and a partial information about the current network state.

Future work will consider the generalization to the case of more than two user classes, and more than two networks coexisting. Our model still applies in these cases, though computations are more difficult. Trying to go towards more realistic scenarios, we could also assume that users have different sets of access networks available, because of the technologies implemented in their terminals, or their geographical location. Finally, it would also be interesting to include user mobility in the model.

\section*{Acknowledgements}

This work has been partially funded by the Bretagne Region through the ARED program, and by the Fondation
Telecom through the ``Futur\&Ruptures'' program.

\bibliographystyle{IEEEtranS}
\bibliography{../BibliographyKaci}

\end{document}